\documentclass[superscriptaddres, twocolumn,notitlepage,showkeys,showpacs,longbibliography]{revtex4-1}
\usepackage[utf8]{inputenc}
\usepackage{stmaryrd}
\usepackage{amsmath}
\usepackage{amssymb}
\usepackage{graphicx}
\usepackage{subfigure}
\usepackage{dcolumn}
\usepackage{bm}
\usepackage{multirow}
\usepackage[colorlinks,linkcolor=blue,citecolor=blue,urlcolor=blue]{hyperref}
\usepackage{color}
\usepackage{ulem}
\usepackage{comment}
\usepackage{booktabs}
\begin{document}
\title{
Quantum oscillations with topological phases in a kagome metal CsTi$_3$Bi$_5$
% application to kagome metal CsTi$_3$Bi$_5$
}
\author{Yongkang Li}
\author{Hengxin Tan}
\author{Binghai Yan}
\email{binghai.yan@weizmann.ac.il}
\affiliation{Department of Condensed Matter Physics, Weizmann Institute of Science, Rehovot 7610001, Israel}
\date{\today}

\begin{abstract}
%Kagome materials $A$V$_3$Sb$_5$ ($A$ = K, Rb, Cs) provide a fertile platform hosting various correlated quantum states such as superconductivity and charge density waves (CDW). Recently, a Ti-based Kagome metal $A$Ti$_3$Bi$_5$ ($A$ = K, Rb, Cs) was synthesized and could serve as a complementary system for investigating pure electronic related phases. Although its band structure has been measured by angle-resolved photoemission spectroscopy (ARPES), the quantum transport evidence for the band topology is still lacking. 
Quantum oscillations can reveal Fermi surfaces and their topology in solids and provide a powerful tool for understanding transport and electronic properties. It is well established that the oscillation frequency maps the Fermi surface area by Onsager's relation. However, the topological phase accumulated along the quantum orbit remains difficult to estimate in calculations, because it includes multiple contributions from the Berry phase, orbital and spin moments, and also becomes gauge-sensitive for degenerate states. In this work, we develop a gauge-independent Wilson loop scheme to evaluate all topological phase contributions and apply it to CsTi$_3$Bi$_5$, an emerging kagome metal.  We find that the spin-orbit coupling dramatically alters the topological phase compared to the spinless case. 
Especially, oscillation phases of representative quantum orbits demonstrate a strong 3D signature despite their cylinder-like Fermi surface geometry. 
Our work reveals the Fermi surface topology of CsTi$_3$Bi$_5$ and paves the way for the theoretical investigation of quantum oscillations in realistic materials.
\end{abstract}

\maketitle

\section{Introduction} \label{introduction}
Kagome lattice, a 2D corner-sharing triangle lattice, has attracted much interest due to its geometric frustration and non-trivial band geometry. Among various materials containing such 2D lattice structure, Kagome material family $A$V$_3$Sb$_5$ ($A$ = K, Rb, Cs)\cite{ortiz2019new} receives special attention since it exhibits many exotic quantum phenomena including $\mathbb{Z}_2$ topology and flat bands\cite{Ortiz2020cs, Ortiz2021superconductivity, hu2022topological}, possible unconventional superconductivity\cite{Ortiz2020cs, Ortiz2021superconductivity, yin2021superconductivity, chen2021double, chen2021roton, tan2021charge, xu2021multiband} and density wave order \cite{chen2021roton, Ortiz2020cs, Ortiz2021superconductivity, yin2021superconductivity, chen2021double, jiang2021unconventional, liang2021three, zhao2021cascade, tan2021charge, park2021electronic, hu2022topological, nie2022charge, li2022emergence}. However, because of the interplay and competition between different correlated states, the origin of these physical properties and their relation to the unique electronic structure remains elusive. 

Recently, a new Ti-based Kagome material $A$Ti$_3$Bi$_5$ ($A$ = K, Rb, Cs) isostructural to $A$V$_3$Sb$_5$ is synthesized\cite{werhahn2022kagome} and investigated\cite{werhahn2022kagome, li2022electronic, yang2022titaniumbased, hu2022nontrivial, yang2022observation, liu2022tunable, jiang2022flat, zhou2023physical, Dong2023CTB}. Unlike V-based $A$V$_3$Sb$_5$ family, the charge density wave (CDW) order is absent in $A$Ti$_3$Bi$_5$ family as shown in transport and scanning tunneling microscopy (STM) experiments\cite{werhahn2022kagome, zhou2023physical, li2022electronic, yang2022superconductivity, hu2022nontrivial, jiang2022flat}. First-principles calculation also shows the absence of lattice structural instability\cite{liu2022tunable}. Hence, $A$Ti$_3$Bi$_5$ could serve as a complementary system to $A$V$_3$Sb$_5$, in which the origin of these exotic phenomena and their relation to electronic properties can be investigated without reference to lattice's effect. For example, the observed two-fold rotational symmetry and orbital selectivity in the electronic structure of $A$Ti$_3$Bi$_5$ \cite{li2022electronic, yang2022superconductivity, hu2022nontrivial, jiang2022flat} may form a pure electronic nematic phase, similar to that in Fe-based high-temperature superconductors\cite{chuang2010nematic}. Understanding the band structure and Fermi surface of $A$Ti$_3$Bi$_5$ is crucial for further investigating these correlating properties.

Quantum oscillation measurement is one way to measure the Fermi surface topology as well as its associated properties like cyclotron mass and carrier mobility\cite{Shoenberg}. More importantly, the phase of the fundamental oscillation is related to the band topology. Usually, a $\pi$ phase shift in the oscillation is regarded as $\pi$ Berry phase which indicates a topological band structure\cite{Mikitik1999Berry, Igor2001Phase, Zhang2005Berry}. The quantum oscillation analysis from this perspective has been carried out in $A$V$_3$Sb$_5$\cite{Shrestha2022CV3Sb5, Fu2021quantum, Broyles2022CV3Sb5, chapai2022magnetic} and also recently in ATi$_3$Bi$_5$\cite{werhahn2022kagome, Dong2023CTB}, which claims nontrivial band topology due to this $\pi$ Berry phase.

The topological phase actually has other contributions entangled with the Berry phase\cite{Aris2018PRX, Aris2018PRB}. Especially in the degenerate case with strong spin-orbit coupling (SOC), such $\pi$ phase may mainly come from orbital or spin magnetic moment other than the Berry phase, as revealed recently in CsV$_3$Sb$_5$ \cite{Heng2023CV3Sb5}. Hence, the analysis of the topological properties based on the phase shift in quantum oscillation should consider all contributions. Apart from the experiment, this phase can be independently evaluated from ab-initio band structures. However, such calculation has to deal with the gauge fixing problem in the presence of degeneracy which is common for centrosymmetric nonmagnetic materials. A numerical study  for all phase contributions without gauge ambiguities has not been explored in detail before.

In this work, we develop a Wilson loop method to determine the quantum oscillation phase and apply it to CsTi$_3$Bi$_5$. We first detail the method which has explicit gauge independence and can be implemented conveniently in the case of degenerate bands. Then combining this method with first-principles calculation we resolve the Fermi surface of CsTi$_3$Bi$_5$ and determine the total oscillation phase for all quantum orbits. Its relation to the Fermi surface geometry and band topology is clarified at last. The 3D nature of several representative quantum orbits present is imprinted in the topological phase, although related Fermi surfaces show a cylinder-like shape. 
Our work provides a useful theoretical tool to investigate the Fermi surfaces and topological electronic properties in materials.

%%%%%%%%%%%%%%%%%%%%%%%%%%%%%%%%%%%%%%%%%%%%%%%%%%%%%%%%%%%%%%%%%%%%%%%%%%%%%%

% Kagome: superconductivity, CDW, PDW, nematicity(rotation symm. breaking), topological band(flat band, Dirac point, VHS), quantum spin liquid(geometry frustration)

% Paper research:

% Kagome synthesize: Ortiz2019new

% unconventional superconductivity: Ortiz2020cs(CVS), Ortiz2021superconductivity(KVS), yin2021superconductivity, chen2021double, chen2021roton, tan2021charge, xu2021multiband, liang2021three, lou2022charge

% Z2 topological insulator: Ortiz2020cs, Ortiz2021superconductivity, hu2022topological

% VHS: kang2022twofold

% Flat bands: hu2022topological

% CDW: Ortiz2020cs, Ortiz2021superconductivity, yin2021superconductivity, chen2021double, jiang2021unconventional, liang2021three, zhao2021cascade, tan2021charge, park2021electronic, hu2022topological, kang2022charge, nie2022charge, li2022emergence, xu2022three, lou2022charge

% PDW: chen2021roton

% rotation symm. breaking: xiang2021twofold, zhao2021cascade, park2021electronic, nie2022charge, li2022emergence, xu2022three

% TR symm. breaking: yang2020giant, jiang2021unconventional, feng2021chiral, mielke2022time, yu2021evidence, xu2022three, khasanov2022time

% AHE: yang2020giant, feng2021chiral

% Last paragraph: It also helps understanding the exotic electronic related phenomena in Kagome metals from its non-trivial topological band structure.

%%%%%%%%%%%%%%%%%%%%%%%%%%%%%%%%%%%%%%%%%%%%%%%%%%%%%%%%%%%%%%%%%%%%%%%%%%%%%

\section{Overview on the quantum oscillation phase} \label{method}

In the presence of a strong magnetic field, the physical quantities (e.g., resistance and magnetization) show oscillation with respect to a magnetic field ($B$) due to the formation of quantized Landau levels (LLs). Under the semiclassical limit in which the scale of $k$-space orbit is much larger than the inverse of magnetic length $l_B^{-1}$ ($l_B=\sqrt{\hbar /eB}$), the oscillation is periodic with respect to $1/B$ and can be expanded as a sum of Fourier series in general:

\begin{equation}
\begin{aligned}
    \delta A = \sum_{i}\sum_{r} A_{i,r} \mathrm{cos}\left[r(l_B^2 S_{F,i}+\theta_{i}+\phi_{M,i}) + \delta_i + \varphi_A \right].
\end{aligned}
\label{decomposition}
\end{equation}

\noindent Here, $A$ is the physical quantity being measured which is usually magnetization $M$ or longitudinal resistivity $\rho_{xx}$, $\delta A$ is the oscillation part and $A_{i,r}$ is the oscillation amplitude for the $r$-th harmonic of the $i$-th extremal orbit. $S_{F,i}$ is the momentum space area of the $i$-th extremal orbit on Fermi surface and determines the $i$-th oscillation frequency. Here the total oscillation phase is decomposed into four parts: $\theta_i$ is the first-order correction to the dynamical phase including the geometry phase and (orbital and spin) magnetic moment phase. $\phi_{M,i}$ is Maslov correction which equals to $\pi$ for a simple closed orbit. $\delta_i$ is dimension related phase resulting from the integration over $k_z$ if a 3D solid is measured (suppose $B$ is along z direction). The last term $\varphi_A$ is measured quantity  ($A$) related phase (see the following discussion).
All phases except $\varphi_A$ depend only on the Fermi surface properties and are universal for any oscillatory quantity. Below we show that each phase can be determined from first-principles calculations to understand experiments. We note that a comprehensive theory on quantum oscillations was established in Refs.~\cite{Aris2018PRX, Aris2018PRB}. We first overview this theory and then introduce the Wilson loop method to compute the topological phase.

\subsection{Phase $\theta$} The first two phases $\theta$ and $\phi_{M}$ (below we focus on a single orbit and ignore the subscript $i$) are related to LLs. In general, there are no simple rules to determine the exact LL for arbitrary band structure. However, in the semiclassical limit, approximate LL can be determined from Bohr-Sommerfeld-like quantization rules. For a group of $D$-fold degenerate bands, the $j$-th LLs can be obtained up to leading order in $l_B^{-1}$ as,
\begin{equation}
    l_{B}^{2} S(E_{a,j}) +\lambda_{a} + \phi_M = 2\pi j + O(l_B^{-2/3}).
\label{Onsager}
\end{equation}
\noindent $a \in \mathbb{Z}_D:=\{1, \ldots, D\}$ is the band index among $D$ degenerate bands and $\lambda_a$ is a  phase that we are interested.  
$\lambda_a$ is equivalent to $\theta$ if there is no degeneracy,i.e., $D=1$. $\phi_{M}$ is Maslov correction and can be determined from the topology of the orbit, which equals $\pi$ for a simple closed orbit.

Because of degeneracy, $D$ LLs create $D$ oscillation terms with the same frequency $F = \hbar S_F/2\pi e$ by Onsager's relation but different phase shift $\lambda_a$. It amounts to a single oscillation term with reduced amplitude $C$ and effective phase shift $\theta$,
\begin{equation}
    \sum^{D}_{a=1} \mathrm{cos}\left[r(l_B^2 S_{F}+\lambda_{a}+\phi_{M})\right] = C\mathrm{cos}\left[r(l_B^2 S_{F}+\theta+\phi_{M})\right].
\label{theta}
\end{equation}
\noindent For example, all bands are doubly degenerate ($D=2$) in the presence of combined inversion and time reversal ($\mathcal{PT}$) symmetries, which is the case of kagome metals CsV$_3$Sb$_5$ and CsTi$_3$Bi$_5$. We regulate $\lambda_{1,2}$ in the range of [$-\pi,\pi$] and then $\mathcal{PT}$ symmetry leads to $\lambda_1 = - \lambda_2 $. Hence, summing two cosine functions in Eq.\eqref{theta} leads to 
\begin{equation}
\begin{aligned}
    \theta &= 
\begin{cases}
    0,& \text{if } |\lambda_1| < \pi/2\\
    \pi, & \text{if } |\lambda_1| > \pi/2\\
\end{cases} \\
    C &= |\mathrm{cos}(\lambda_1)| &, \\    
\end{aligned}
\label{PT_theta}
\end{equation}
One can find $\theta$ is a quantized topological invariant (0 or $\pi$) \cite{Aris2018PRX} insensitive to orbit details.

In general, phase $\lambda_a$ can be determined from the spectrum $\{e^{i\lambda_a}\}_{a=1}^{D}$ of propagator\cite{Aris2018PRX,Aris2018PRB} 
\begin{equation}
\mathcal{A}[\mathfrak{o}]=\overline{\exp}\left[i \oint_\mathfrak{o}\left\{(\boldsymbol{\mathrm{A}}+\boldsymbol{\mathrm{R}}) \cdot d \boldsymbol{k}+Z\left(\sigma^z / v^{\perp}\right) d k\right\}\right].
\label{propagator}
\end{equation}

\noindent Here $\overline{\exp}$ means path-ordered product, $\boldsymbol{\mathrm{A}}(\boldsymbol{k})_{m n}=i\left\langle u_{m \boldsymbol{k}}| \nabla_{\boldsymbol{k}} u_{n \boldsymbol{k}}\right\rangle$ is non-Abelian Berry connection and
\begin{align}
    \boldsymbol{\mathrm{R}}_{m n} \cdot d \boldsymbol{k}&=\sum_{l \notin \mathbb{Z}_D} \mathrm{A}_{m l}^x \Pi_{l n}^y d k_x / 2 v_y+(x \leftrightarrow y) \notag\\
    &=-i\hbar\sum_{l \notin \mathbb{Z}_D} \frac{\Pi_{m l}^x \Pi_{l n}^y}{\varepsilon_{m \boldsymbol{k}}-\varepsilon_{l \boldsymbol{k}}} \frac{d k_x}{2 v_y} + (x \leftrightarrow y) \label{Roth form exp1} \\
    &=-(M_z/ev^{\perp})dk,
\end{align}
\noindent is Roth term and represents the orbital correction ($-M_z B_z$) to the band energy. $\boldsymbol{\Pi}(\boldsymbol{k})_{l n}=\left\langle u_{l \boldsymbol{k}}|(1/\hbar)\nabla_{\boldsymbol{k}}\hat{H}(\boldsymbol{k})| u_{n \boldsymbol{k}}\right\rangle$ is velocity matrix element and $\boldsymbol{v}=\boldsymbol{\Pi}_{n n}$ is group velocity. $\epsilon_{m\boldsymbol{k}}$ is band energy and $v^{\perp}$ is the velocity in $xy$ plane. $M_{z}=i(e\hbar/2)\sum_{l \notin \mathbb{Z}_D} \Pi_{m l}^{x} \Pi_{l n}^{y}/(\varepsilon_{m \boldsymbol{k}}-\varepsilon_{l \boldsymbol{k}}) - (x \leftrightarrow y)$ is the self rotation part of orbital magnetic moment\cite{Xiao2010}. Furthermore,
$\sigma_{z,mn} = \left\langle u_{l \boldsymbol{k}}|\hat{\sigma}_z| u_{n \boldsymbol{k}}\right\rangle$ ($\hat{\sigma}_z$ is spin Pauli matrix) and $Z=g_0 \hbar/4m$. The last term is the spin Zeeman term. Once the propagator ($\mathcal{A[\mathfrak{o}]}$) is known, the phase $\lambda_a$ can be easily obtained by diagonalizing it.

Though its formulation is clear in theory, the numerical calculation of this propagator needs to deal with the derivatives in the Berry connection. Besides, the multi-band magnetic moment (including orbital and spin) is a gauge covariant quantity whose matrix elements depend on the gauge. If a random gauge is chosen, the magnetic moment transforms independently at each point along the orbit, rendering the \eqref{propagator} meaningless. To deal with these problems, one can choose a smooth gauge by finding the maximally localized Wannier function\cite{Vanderbilt1997}. Alternatively, the Wilson loop method\cite{Fukui2005C1, Yu2021Z2, Vanderbilt2011PRB} can be applied to avoid the choice of any specific gauge. Below, we shall use the Wilson loop method for the calculation of $\lambda_a$.

\begin{figure}
\centering
\includegraphics[width=0.8\linewidth]{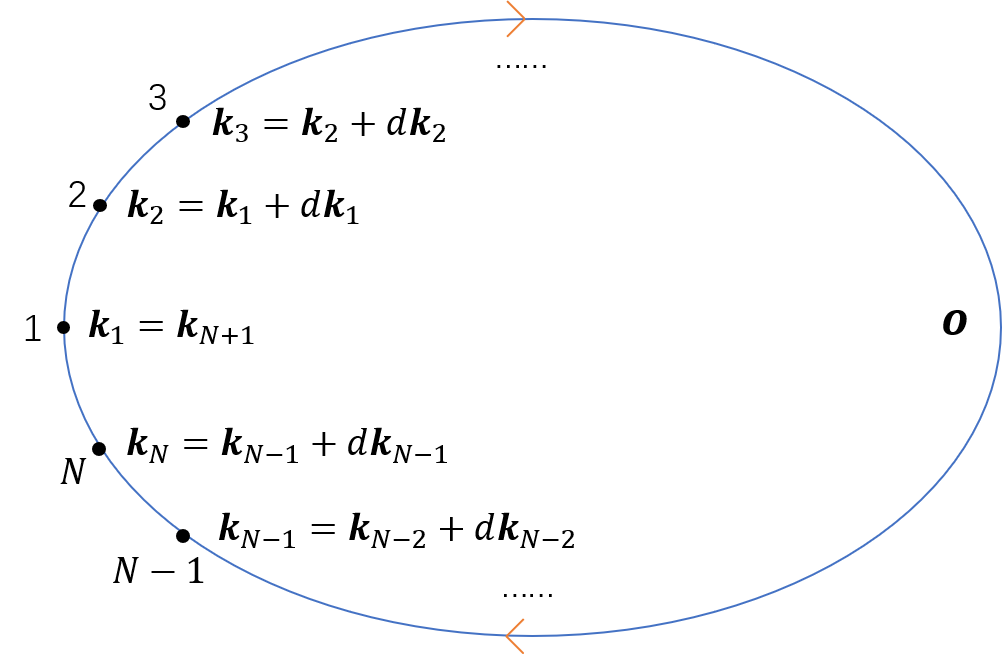}
\caption{\label{Fig1}
The Wilson loop $\mathfrak{o}$ for calculation of propagator $\mathcal{A}$. Here it's discretized to N points and the circulation direction is clockwise.
}
\end{figure}

In this way, the quantum orbit is discretized into $N$ segments (Fig.\ref{Fig1}) and the propagator is written as the product for each segment. If the segment is small enough, the exponent can be split into Berry connection and magnetic moment parts.

\begin{equation}
\begin{aligned}
    &\mathcal{A}[\mathfrak{o}] = \prod_{i=1}^{N} \mathrm{exp}\left\{i\left[(\boldsymbol{\mathrm{A}}(\boldsymbol{k_i}) + \boldsymbol{\mathrm{R}}(\boldsymbol{k}_i))\cdot d\boldsymbol{k_i}+Z\frac{\sigma^{z}}{v^{\perp}}|d\boldsymbol{k}_i|\right]\right\} \\
    & \approx \prod_{i=1}^{N}  \mathrm{exp}\left[i\boldsymbol{\mathrm{A}}(\boldsymbol{k_i}) \cdot d\boldsymbol{k_i}\right] \mathrm{exp}\left[i\boldsymbol{\mathrm{R}}(\boldsymbol{k}_i) \cdot d\boldsymbol{k_i}+i Z\frac{\sigma^{z}}{v^{\perp}}|d\boldsymbol{k}_i|\right].
\label{discretization}
\end{aligned}
\end{equation}

\noindent For numerical calculation, the Berry connection part is usually expressed by an overlap matrix $M^{i} =\mathrm{exp}\left[i\boldsymbol{\mathrm{A}}(\boldsymbol{k_i}) \cdot d\boldsymbol{k_i}\right] $. $M_i$ is a $D$ by $D$ matrix with  $M^{i}_{mn}=\left\langle u_{m \boldsymbol{k_{i+1}}}|u_{n \boldsymbol{k_i}}\right\rangle$.

% \begin{equation}
% \begin{aligned}
%     \mathrm{exp}\left[i\boldsymbol{\mathrm{A}}(\boldsymbol{k_i}) \cdot d\boldsymbol{k_i}\right] = M^{i}.
% \label{overlap matrix}
% \end{aligned}
% \end{equation}

% \noindent here $| u_{m\boldsymbol{k}_{i}} \rangle$ is the m-th Bloch states in D degenerate space and $M_i$ is a D by D matrix.

The last ingredient for the propagator is an appropriate expression for the Roth term which shows explicit gauge covariance. It can be written as a summation of velocity matrix elements over \textit{all} other states as in \eqref{Roth form exp1}. 
Instead, we propose another method that considers only $D$ degenerate states on the Fermi surface using the covariant derivative\cite{Ceresoli2006PRB}. The covariant derivative is defined as
\begin{align}
    \left|D_\alpha u_{n \boldsymbol{k}}\right\rangle &=Q_{\boldsymbol{k}}\left|\partial_\alpha u_{n \boldsymbol{k}}\right\rangle, \quad
    Q_{\boldsymbol{k}} := I-\sum_{a \in \mathbb{Z}_D} \left|u_{a \boldsymbol{k}}\right\rangle \left\langle u_{a \boldsymbol{k}}\right|.
\label{covariant derivative}
\end{align}
In numerical calculation, it can be evaluated as an appropriate finite difference\cite{Ceresoli2006PRB}
\begin{equation}
\left|D_\alpha u_{n \boldsymbol{k}}\right\rangle=\frac{1}{2|\boldsymbol{q}_\alpha|}\left(\left|\overline{u}_{n, \boldsymbol{k}+\boldsymbol{q}_\alpha}\right\rangle-\left|\overline{u}_{n, \boldsymbol{k}-\boldsymbol{q}_\alpha}\right\rangle\right),
\label{finite difference}
\end{equation}
\noindent where the dual state $\left|\overline{u}_{n, \boldsymbol{k}+\boldsymbol{q}}\right\rangle$ is a linear combination of $\left|u_{n, \boldsymbol{k}+\boldsymbol{q}}\right\rangle$ and has the property $\left\langle \overline{u}_{m \boldsymbol{k}}| \overline{u}_{n \boldsymbol{k}+\boldsymbol{q}}\right\rangle=\delta_{m n}$. This ensures the orthogonality between the covariant derivative and states in degenerate space, i.e. $\left\langle u_{m \boldsymbol{k}}| D_\alpha u_{n \boldsymbol{k}}\right\rangle=0$. Dual states are constructed as 
\begin{equation}
\left|\overline{u}_{n, \boldsymbol{k}+\boldsymbol{q}}\right\rangle=\sum_{n^{\prime}}\left(S_{\boldsymbol{k}, \boldsymbol{k}+\boldsymbol{q}}^{-1}\right)_{n^{\prime} n}\left|u_{n^{\prime}, \boldsymbol{k}+\boldsymbol{q}}\right\rangle
\label{dual states}
\end{equation}

\noindent and
\begin{equation}
\left(S_{\boldsymbol{k}, \boldsymbol{k}+\boldsymbol{q}}\right)_{n n^{\prime}}=\left\langle u_{n \boldsymbol{k}} | u_{n^{\prime}, \boldsymbol{k}+\boldsymbol{q}}\right\rangle .
\end{equation}

Using covariant derivative, Eq. \eqref{Roth form exp1} is expressed only by states inside the degenerate space
\begin{equation}
\begin{aligned}
    &\boldsymbol{\mathfrak{\mathrm{R}}}_{m n} \cdot d \boldsymbol{k} = -\frac{i}{\hbar}\sum_{l \notin \mathbb{Z}_D} \mathrm{A}_{m l}^x (\varepsilon_{n \boldsymbol{k}}-\varepsilon_{l \boldsymbol{k}}) \mathrm{A}_{l n}^y d k_x / 2 v_y+(x \leftrightarrow y) \\
    &= -\frac{i}{\hbar}\sum_{l \notin \mathbb{Z}_D} \left\langle D_{x} u_{m \boldsymbol{k}}| u_{l \boldsymbol{k}}\right\rangle (\varepsilon_{n \boldsymbol{k}}-\varepsilon_{l \boldsymbol{k}}) \left\langle u_{l \boldsymbol{k}}| D_{y} u_{n \boldsymbol{k}}\right\rangle d k_x / 2 v_y \\ 
    & \qquad + (x \leftrightarrow y) \\
    &= -\frac{i}{\hbar}\left\langle D_{x} u_{m \boldsymbol{k}}| \varepsilon_{n \boldsymbol{k}}-\hat{H}(\boldsymbol{k}) | D_{y} u_{n \boldsymbol{k}}\right\rangle d k_x / 2 v_y+(x \leftrightarrow y).
\end{aligned}
\label{Roth form exp2}
\end{equation}

\noindent In Appendix we show both Eq. \eqref{Roth form exp1} and Eq. \eqref{Roth form exp2} are gauge independent, which can be implemented easily in first-principles calculation. The Eq. \eqref{Roth form exp1} is practical for tight-binding models with a small number of bands but quite tedious if the total number of bands is large. The Eq. \eqref{Roth form exp2} avoids these problems and focuses only on the degenerate space and it is convenient when covariant derivatives can be easily calculated.

% Besides it has the drawback that the Roth one-form should only depend on the degenerate space being considered but \eqref{Roth form exp1} involves other states and hinders this property.

% To calculate the Roth one-form, two covariant derivatives in the orthogonal direction and hence four neighboring states are needed at each k point. Even if it avoids the tedious summation over other states, finding the neighboring states usually involves additional diagonalization on k points outside the orbit. Unless the states of degenerate space are previously given at a neighborhood of the orbit, this way may be more expensive than the first.

\subsection{Phase $\delta$} 
\begin{figure} %[htbp]
\centering
\subfigure[\label{Fig2-sphere}]
{
\includegraphics[width=0.3\linewidth]{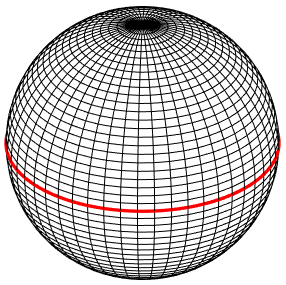}
}
\subfigure[\label{Fig2-cylinder1}]
{
\includegraphics[width=0.3\linewidth]{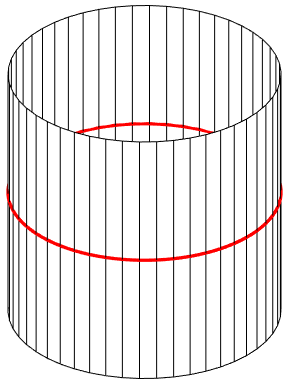}
}
\subfigure[\label{Fig2-cylinder2}]
{
\includegraphics[width=0.3\linewidth]{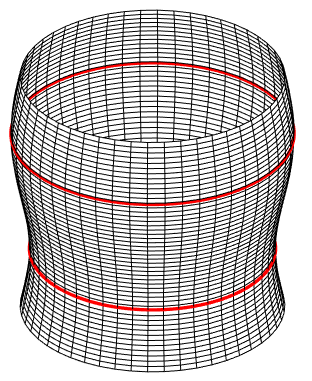}
}
\caption{\label{Fig2}
(a) Spherical Fermi surface, (b) perfect cylindrical Fermi surface, and (c) nearly cylindrical Fermi surface. In both cases, the red circle shows the extremal orbit.
}
\end{figure}

{The above discussion about phase $\theta$ is for a single $k$-plane perpendicular to the magnetic field. For 3D material, one needs to integrate over $k_z$ to get the contribution from the whole Fermi surface. Extremal orbits will dominate in the integration and this procedure will introduce another phase $\delta$ for each of them, which is generally $\pm \pi/4$ ($+$ for minimum cross-section and $-$ for maximum cross-section). $\delta = 0$ for 2D material since there is only one k-plane.} But for a nearly cylindrical Fermi surface (e.g., Fig.\ref{Fig2-cylinder2}), $\delta$ lies between these two limits.  Below we adopt a simple model from Refs.~\cite{Shoenberg, Shoenberg1973AnomalousAA} to determine $\delta$ for every extremal orbit that lies in a mirror plane. Here, we assume $\mathcal{PT}$ symmetry for simplicity.

The oscillation of 3D Fermi surface is calculated first for a 2D plane with thickness $dk_z$ and then integrate with respect to $k_z$, i.e.

\begin{equation}
\begin{aligned}
    A_r &= \sum_a \int dk_z \, A_r(k_z) \mathrm{cos}\left[r (2\pi\frac{F(k_z)}{B} + \lambda_a(k_z)) + \phi_M \right] \\
    & \propto \int dk_z \, A_r(k_z) \mathrm{cos}\left[r (2\pi\frac{F(k_z)}{B} + \theta(k_z)) + \phi_M \right],
\end{aligned}
\label{superposition z}
\end{equation}

\noindent where $A_r(k_z)$ is the oscillation amplitude of 2D plane, which depends on $k_z$ through cyclotron frequency $F(k_z)$ and cyclotron mass $m(k_z)$. The relative change of $F(k_z)$ and $m(k_z)$ in the interval where the integral is appreciable is usually small. Hence in the integration of Eq. \eqref{superposition z}, $A_r(k_z)$ can be treated approximately as a constant while $F(k_z)$ in the cosine function can't be treated as fixed because $F(k_z)\gg B$. Maslov phase $\phi_M$ remains constant as long as the orbit on the Fermi surface doesn't change its topology. Moreover, $\mathcal{PT}$ symmetry cause the phase $\theta(k_z)$ quantized to 0 or $\pi$ as in Eq. \eqref{PT_theta}. So the $k_z$ dependence of $\theta$ can also be ignored and only the $k_z$-variation of $F(k_z)$ needs to be considered.

We expand $F(k_z)$ near its extremal value to the fourth order and all odd orders are zero due to mirror symmetry.

\begin{equation}
\begin{aligned}
    F(k_z) = F_0 + \frac{1}{2}F_{2}k_z^2 + \frac{1}{24}F_{4}k_z^4.
\end{aligned}
\label{expansion F}
\end{equation}

\noindent Introducing dimensionless variable $x = (2r|F_{2}|/B)^{1/2} k_z$ and $\alpha = \mathrm{sgn}(F_{2})\frac{F_{4}B}{24 r |F_{2}|^2}$ then the integration can be calculated as

\begin{widetext}
\begin{eqnarray}
    \begin{aligned}
    A_r &\propto \mathrm{Re} \int\mathrm{exp}\left[i(2\pi r\frac{F(k_z)}{B} + r\theta + \phi_M)\right] dk_z \\
    &\propto \mathrm{Re} \ \mathrm{exp}\left[i(2\pi r\frac{F_0}{B} + r\theta + \phi_M) \right] \int \mathrm{exp} \left[\mathrm{sgn}(F_{2})i\frac{\pi}{2} x^2 (1+\alpha x^2)\right]dx \\
    &\propto \mathrm{cos}\left[r (2\pi\frac{F_0}{B} + \theta) + \phi_M + \delta \right].
\end{aligned}
\end{eqnarray}
\label{integral u}
\end{widetext}

\noindent where phase $\delta$ is the argument of the last integral
\begin{equation}
\begin{aligned}
    \delta &= \mathrm{arg} \left\{\int^{x_m}_{x_m}\mathrm{exp}\left[\mathrm{sgn}(F_{2})i\frac{\pi}{2} x^2 (1+\alpha x^2)\right] dx\right\}.
\end{aligned}
\label{delta}
\end{equation}
$\delta$ was numerically determined by carrying out the integral with given value $\alpha$ \cite{Shoenberg, Shoenberg1973AnomalousAA}, for which $F_{2}$, $F_{4}$ can be found from the polynomial fitting of $F(k_z)$ around the extremal orbit. The integral limit $x_m$ can be taken as $\infty$ when $\alpha>0$ because the main contribution comes from $x\approx 0$.
However, this argument does not apply when $\alpha<0$ due to the two extra artificial extrema. Since the real cross-section varies monotonically on either side of $x=0$, $x_m$ should be taken less than the turning point $1/\sqrt{2|\alpha|}$ to avoid these artificial extrema. In calculation, the argument of the integral goes to a steady value before the turning point, which should be assigned as $\delta$. It's obvious that $\delta= 0$ from Eq.~\eqref{superposition z} when $F(k_z)=F_0$.
For a general 3D material, if $\alpha \rightarrow 0$ (i.e., $F_4 \rightarrow 0$ and $F_2 k_z^2$ is the leading dispersion), one can get $\delta=\pm \pi/4$. Otherwise, $\delta$ may take a value between 0 and $\pm \pi/4$.

\subsection{Phase $\varphi_A$}
The last phase $\varphi_A$ depends on the type of physical quantity $A$. When $A$ is the density of states (DOS), this phase vanishes $\varphi_{DOS}=0$. For other quantities, $\varphi_A$ represents the connection between the oscillation of $A$ and the oscillation of DOS. For example, $\varphi_M=\pi/2$ if $A$ is sample magnetization, and $\varphi_{\chi}=\pi$ if $A$ is magnetic susceptibility. In four terminal devices, the longitudinal conductivity $\sigma_{xx}$ oscillates in phase with DOS hence $\varphi_{\sigma}=0$. But since $\sigma_{xx}=\rho_{xx}/(\rho_{xx}^2+\rho_{xy}^2)$, the resistivity $\rho_{xx}$ can be in phase (if $\rho_{xx}\ll\rho_{xy}$) or out of phase (if $\rho_{xx}\gg\rho_{xy}$) with $\sigma_{xx}$, so $\varphi_{\rho} = 0$ if $\rho_{xx}\ll\rho_{xy}$ or $\varphi_{\rho} = \pi$ if $\rho_{xx}\gg\rho_{xy}$\cite{Fei2015BiTeCl, Fu2021quantum}.

To summarize, all the phases in the oscillation term Eq. \eqref{decomposition} have the following intuitive explanations. First, the magnetic-field-dependent term $l_B^2 S_F$ is given by the combination of the de Broglie phase (determined by the number of wavelengths in an orbit) and the Aharonov–Bohm phase. Then there is a phase $\lambda_a$ associated with each orbit and each band coming from geometric effects and magnetic moment energy. $\lambda_a$ of degenerate bands for the same orbit will combine to give the phase $\theta$. The reflection of the wave packet at turning points in the orbit causes phase $\phi_M$. These phases are the total phase for a single orbit lying in the $kx-ky$ plane. For 3D materials, $k_z$ integration needs to be carried out to incorporate the whole Fermi surface's contribution, which gives phase $\delta$. At last, depending on what quantity $A$ is measured, there will be another phase $\phi_A$ if the oscillation of $A$ is not synchronized with the oscillation of DOS.

\section{Results and Discussions}

The crystal structure of CsTi$_3$Bi$_5$ is fully relaxed within the Density Functional Theory (DFT) as implemented in the Vennia $ab$-$inito$ Simulation Package \cite{PRB54p11169, VASP}. The cutoff energy for the plane-wave basis set is 300 eV. The force convergence criteria is 5 meV/\AA.
The electronic structure is calculated with the full-potential local-orbital minimum-basis code (FPLO) \cite{FPLO}.
The default atomic basis sets are employed for the wave function expansion.
The generalized gradient approximation parameterized by Perdew, Burke, and Ernzerhof (PBE) \cite{PBE} is employed to mimic the exchange-correlation interaction between electrons throughout.
The Brillouin zone is sampled by a $k$-mesh of 12$\times$12$\times$6.
The tight-binding Hamiltonian of CsTi$_3$Bi$_5$ is extracted via the maximally localized Wannier functions \cite{Vanderbilt1997} as implemented in FPLO, which enforces all crystal symmetries.
The Wannier basis set is composed of the Ti $d$ and Bi $p$ orbitals.
The Fermi surface is calculated with the tight-binding Hamiltonian on a $k$-mesh of $300\times300\times100$.

We mention that the above Wilson loop method for the total oscillation phase shift has been successfully applied to the $\mathcal{PT}$ symmetric kagome metal CsV$_3$Sb$_5$ \cite{Heng2023CV3Sb5}, which predicted consistent results with experiments.
In the following, we will apply the Wilson loop method to the recently discovered kagome superconductor CsTi$_3$Bi$_5$ \cite{werhahn2022kagome} to further demonstrate the reliability of this method.
We note here that the characterization of the dimensionality of the quantum orbit by the phase $\delta$ has not been discussed in our previous work on CsV$_3$Sb$_5$.

\begin{figure}[htbp]
\centering
\includegraphics[width=1\linewidth]{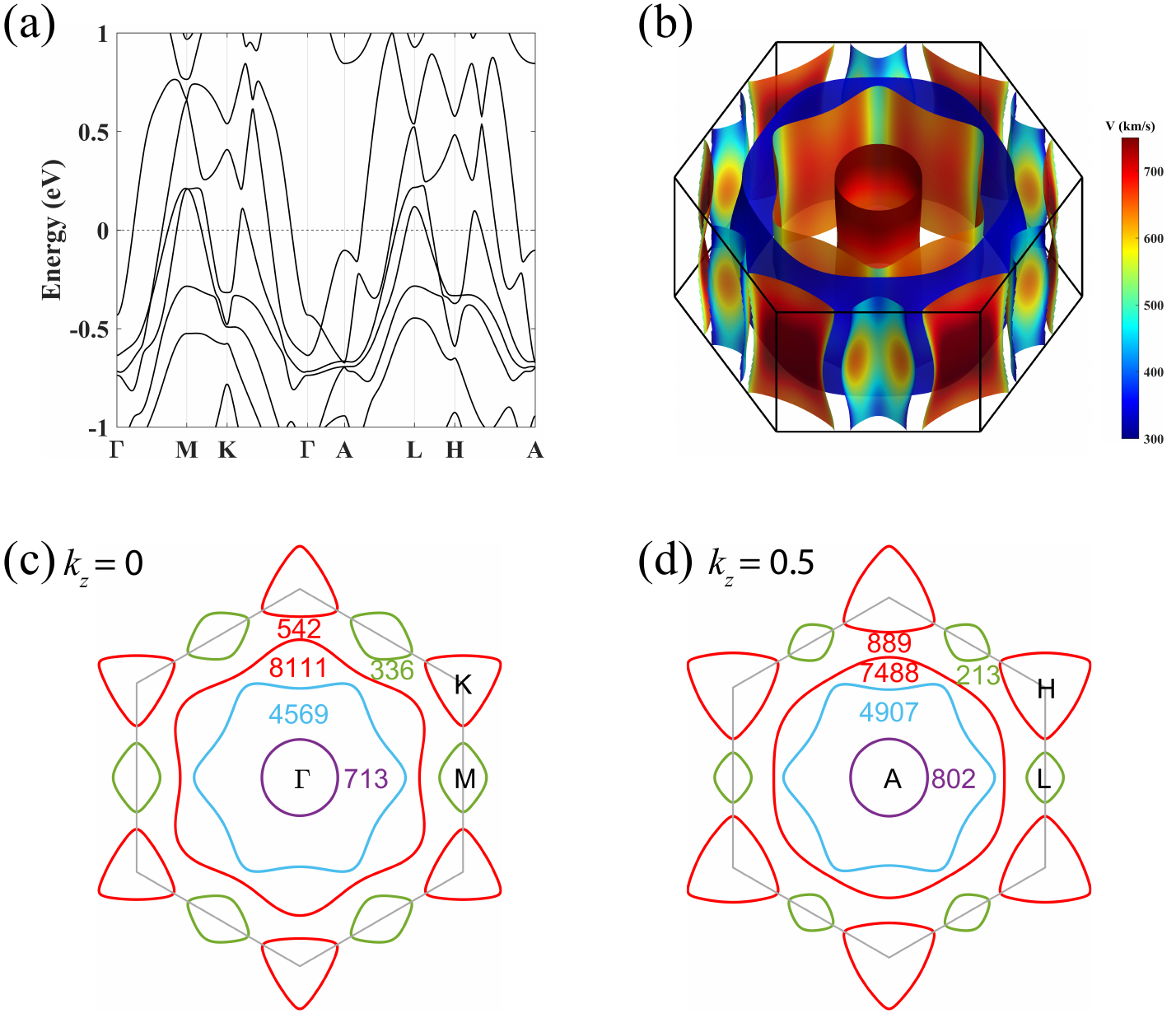}
\caption{\label{Fig3}
(a) Band structure of CsTi$_3$Bi$_5$ with SOC. (b) 3D Fermi surface of CsTi$_3$Bi$_5$, where the color representing the Fermi velocity is used to distinguish between different Fermi surfaces. (c) Fermi surfaces in $k_z=0$ and (d) $k_z=0.5$ (in units of $2\pi/c$, $c$ is the lattice constant) mirror plane at the Fermi energy. The grey hexagon is the first Brillouin zone. Fermi surfaces with the same color come from the same band. The cyclotron frequencies (in units of T) are given.
}
\end{figure}

The band structure of CsTi$_3$Bi$_5$ with spin-orbit coupling is plotted in Fig.\ref{Fig3}(a), which contains rich topological properties.
Due to the $\mathcal{PT}$ symmetry in CsTi$_3$Bi$_5$, each band is doubly degenerate.
Characteristic features of the kagome lattice, such as Dirac points at K/H points away from the Fermi level which are gapped by SOC, van Hove singularities at M/L, and flat bands along M-K/L-H lines \cite{hu2022nontrivial, yang2022observation, jiang2022flat}, are shown. There are also type II Dirac crossings on the $\Gamma$-M and A-L lines, which form a Dirac nodal line \cite{hu2022nontrivial, yang2022observation, jiang2022flat} in the $\Gamma$-M-A plane. Besides, both the experiment and theory have shown that CsTi$_3$Bi$_5$ has topological Dirac surface states at the $\overline{\Gamma}$ point on the (001) surface \cite{yang2022titaniumbased,yang2022observation,jiang2022flat,hu2022nontrivial,zhou2023physical}.

% \begin{figure*}[htbp]
% \centering
% \begin{minipage}{0.3\linewidth}
%     \subfigure[\label{Fig3-band-noSOC}]{
%     \includegraphics[width=1\linewidth]{Fig3-band_structure_noSOC.png}
%     }
    
%     \subfigure[\label{Fig3-band-SOC}]{
%     \includegraphics[width=1\linewidth]{Fig3-band_structure_SOC.png}
%     }
% \end{minipage}
% \begin{minipage}{0.3\linewidth}
%     \subfigure[0meV,$k_z=0$\label{Fig3-FS_kz=0}]{
%     \includegraphics[width=0.7\linewidth]{Fig3-FS_0meV_kz0.png}
%     }

%     \subfigure[0meV,$k_z=0.5$\label{Fig3-FS_kz=0.5}]{
%     \includegraphics[width=0.7\linewidth]{Fig3-FS_0meV_kz0.5.png}
%     }
% \end{minipage}
% \begin{minipage}{0.3\linewidth}
%     \subfigure[\label{Fig3-FS3D}]{
%     \includegraphics[width=1\linewidth]{Fig3-FS3D.png}
%     }
% \end{minipage}
% \caption{\label{Fig3}
% (a) Band structure without or (b) with SOC of CsTi$_3$Bi$_5$. (c) Fermi orbits in $k_z=0$ mirror plane and (d) $k_z=0.5$ (in units of $2\pi/c$, $c$ is the lattice constant) mirror plane at Fermi energy $E_F=0$meV. The grey hexagon is the boundary of the first Brillouin zone. Different color represents the orbits coming from different bands. The cyclotron frequencies (in units of T) are also plotted in the figure. (e) 3D Fermi surface with SOC of CsTi$_3$Bi$_5$. The color map represents the magnitude of Fermi velocity.
% }
% \end{figure*}

The band structure on the $k_z = 0$ plane looks similar to the band structure on the $k_z = 0.5$ plane (in units of 2$\pi$/$c$, $c$ is the lattice constant), which indicates the quasi-two-dimensional feature of the electronic structure of CsTi$_3$Bi$_5$. Indeed, the 3D Fermi surface shown in Fig.\ref{Fig3}(b) shows a good cylindrical shape for all pieces.
There are totally four bands crossing the Fermi level creating five pieces of the Fermi surface. By sweeping $k_z$, all extremal quantum orbits perpendicular to the $z$-direction are found to locate at the two mirror planes $k_z=0$ and $k_z=0.5$, shown in Fig.\ref{Fig3}(c) and (d). The initial experiment reported an oscillation frequency of 200 T \cite{werhahn2022kagome}.
A more recent transport experiment \cite{Dong2023CTB} reported a series of oscillation frequencies, ranging from 217 to 1013 T.
Our calculations show agreement with the experiments in the low-frequency region. For example, the calculated frequencies of 213, 336, and 542 T might correspond to the observed frequencies of 200/217, 281, 498 or 594 T, respectively.
We notice that our calculated frequencies are slightly different from the calculations in Ref. \onlinecite{Dong2023CTB}, which might be induced by the mismatch of Fermi energy and/or different calculation parameters employed.

The cyclotron masses $m^*$ of all calculated quantum orbits are summarized in Table \ref{Table_CsTi3Bi5}.
Except for the two small pockets (336 and 213 T) around M/L points, all other orbits are electron pockets, whose cyclotron masses are defined as positive.
The two largest hexagonal orbits centered around the $\Gamma$ point (7488 and 8111 T) have the largest cyclotron masses (1.6$\sim$1.7) while others have relatively small cyclotron masses.

The different quantum oscillation phases, as mentioned above, of all orbits are calculated and listed in Table.\ref{Table_CsTi3Bi5}.
Here every cyclotron orbit is a simple closed curve; thus the Maslov correction $\phi_M=\pi$ is omitted in the table.
The phase $\lambda_a$ is calculated by Eq. \eqref{Roth form exp2} with random gauge choices to test the gauge invariance, which presents the same results.
We also confirm the relation $\lambda_1 = -\lambda_2$ for any two degenerate quantum orbits imposed by the $\mathcal{PT}$ symmetry.
Thus only the positive one $\lambda_1$ is listed. The Berry phases without ($\phi_{B0}$) or with SOC ($\phi_{B}$) are also listed for comparison.
According to our previous discussion of Eq. \eqref{PT_theta}, the final phase shift of the quantum orbit $\theta$ must be quantized to either 0 or $\pi$, depending on the magnitude of $\lambda_1$, as listed in Table.\ref{Table_CsTi3Bi5}.
From these phases, it's clear that phase $\lambda_1$ is in general different from the Berry phase $\phi_B$ due to the orbital and spin magnetic moment contribution. Also for the $\mathcal{PT}$-symmetric system, the topology of the quantum orbit is not equivalent to the band topology of the individual Fermi surface. For example, the quantum orbits of 336 T (around M) and 8111 T (around $\Gamma$) have Berry phases $\phi_B$ close to 0 but the oscillation phase shifts are $\pi$.
On the contrary, the quantum orbit of 4907 T (around $A$) has a Berry phase close to $\pi$ but a zero oscillation phase shift.
We note here that the strong SOC is important because these orbits have only a trivial Berry phase in the spinless case.
Therefore, the incorporation of the magnetic moment contribution in the oscillation phase by SOC is crucial and the quantum phase shift extracted from the Landau fan diagram should be interpreted more carefully, rather than just interpreting it as the Berry phase.
The recent experiment \cite{Dong2023CTB} finds that the quantum orbit of 281 T is non-trivial with a $\pi$ phase shift ($\theta=\pi$), which is consistent with our calculated non-trivial quantum orbit of 336 T.

Because the 3D Fermi surface is nearly cylindrical, the dimension-related phase $\delta$ should be determined by considering higher order terms in the expansion of $F(k_z)$ in Eq. \ref{expansion F}. From numerical calculations, the frequency $F$ and cyclotron mass $m^*$ of all extremal orbits have a small relative change on the Fermi surface (less than $5\%$ in the interval $|\Delta k_z| \leq 0.1$).
Since CsTi$_3$Bi$_5$ has $\mathcal{PT}$ symmetry and all extremal orbits locate in mirror planes, Eq. \eqref{delta} applies, which is used to calculate phase $\delta$.
The phase $\delta$ is calculated with the magnetic field $B$ varying from 5 T to 40 T, covering the range of $B$ in general oscillation experiments \cite{Fu2021quantum, Shrestha2022CV3Sb5, Broyles2022CV3Sb5}.
The variation of $\delta$ is very small in the considered $B$ range. Thus the $\delta$ can be approximately treated as a constant, whose average value is listed in Table \ref{Table_CsTi3Bi5}. It shows that all quantum orbits except for the 213 and 802 T ones have a phase $\delta$ quite close to $\pm \pi/4$.
Therefore, most orbits should be classified as 3D cases in quantum oscillation, even though the Fermi surfaces in Fig. \ref{Fig3}(b) show a strong quasi-2D feature.
On the other hand, the Fermi surface around A is almost dispersionless along $k_z$, so the $\delta$ for the quantum orbit of 802 T is closer to zero than others. As a result, this quantum orbit is 2D. However, the quantum orbit of 713 T which comes from the same Fermi surface as the 802 T orbit but on the $k_z=0$ plane, has a $\delta=\pi/4$.
Consequently, the character (2D or 3D) of a quantum orbit should not be simply determined from the appearance of the related Fermi surface in the 3D $k$ space.

\begin{table}
\centering
\caption{\label{Table_CsTi3Bi5} Extremal orbits of Fermi surfaces of CsTi$_3$Bi$_5$ at Fermi energy. Frequency (Freq.) is in units of T. $k_z$ refers to the $k_z$ plane (in units of $2\pi/c$, $c$ is the lattice constant) where the corresponding extremal orbit is located in. The underlined frequency indicates a minimal Fermi surface cross-section and the others correspond to a maximal cross-section.
The cyclotron mass $m^*$ is in units of bare electron mass $m_0$, where positive and negative values are for electron and hole pockets respectively. All orbits have Maslov correction $\phi_M=\pi$. $\phi_{B0}$ ($\phi_B$) is Berry phase without (with) SOC and $\lambda_1$ is the phase of one of the band between two degenerate bands, defined in Eq. \eqref{propagator}. $\delta$ is the phase related to the Fermi surface dimensionality. All phases are in units of $\pi$.
}
\renewcommand\arraystretch{1.1}
\begin{ruledtabular}
\begin{tabular}{rccccccc}
    Freq. & $k_z$ & $m^*$  & $\phi_{B0}$ & $\phi_B$ & $\lambda_1$ & $\theta$ & $\delta$ \\
    (T) & ($2\pi/c$) & ($m_0$)   & ($\pi$) & ($\pi$) &   ($\pi$) & ($\pi$) & ($\pi$)  \\
    \cline{1-8}
    \uline{213}  & 0.5  & $-$0.24  & 0 & 0.08 & 0.40 & 0 & 0.22 \\
    336          & 0    & $-$0.22  & 0 & 0.16 & 0.59 & 1 & -0.25\\
    \uline{542}  & 0    & 0.22     & 0 & 0.50 & 0.23 & 0 & 0.25 \\
    \uline{713}  & 0    & 0.26     & 0 & 0.30 & 0.33 & 0 & 0.25 \\
    802          & 0.5  & 0.24     & 0 & 0.33 & 0.12 & 0 & -0.14 \\
    889          & 0.5  & 0.32     & 1 & 0.26 & 0.10 & 0 & -0.25 \\
    \uline{4569} & 0    & 0.72     & 0 & 0.74 & 0.42 & 0 & 0.25 \\
    4907         & 0.5  & 0.78     & 0 & 0.92 & 0.22 & 0 & -0.25 \\
    \uline{7488} & 0.5  & 1.62     & 0 & 0.49 & 0.81 & 1 & 0.25 \\
    8111         & 0    & 1.68     & 0 & 0.38 & 0.82 & 1 & -0.25 \\
\end{tabular}
\end{ruledtabular}
\end{table}

\section{Conclusion}
We theoretically studied the quantum oscillations by revealing their frequencies and topological phases through a Wilson loop method in CsTi$_3$Bi$_5$. 
We revealed three quantum orbits with $\theta = \pi$ phase shift. 
Despite most Fermi surfaces are quasi-2D, 
the dimensional-related phase $\delta$, beyond the angle-dependent frequency, clearly indicates their 3D nature. 
Our method can be applied to other quantum materials and provides a general way to study quantum oscillations assisted by first-principles calculations.

\section*{Acknowledgement} 
B.Y. acknowledges the financial support by the European Research Council (ERC Consolidator Grant ``NonlinearTopo'', No. 815869) and the ISF - Personal Research Grant	(No. 2932/21).

\section*{Appendix}
The most general gauge transformation is a $U(D)$ basis transformation among the degenerate bands
\begin{align}
    \left|u_{n \boldsymbol{k}}\right\rangle &\rightarrow \sum_{m=1}^D U(\boldsymbol{k})_{m n}\left|u_{m \boldsymbol{k}}\right\rangle , \quad U^{-1}=U^{\dagger}, 
    \label{gauge transformation}
\end{align}

It has already been shown that the propagator $\mathcal{A}[\mathfrak{o}]$ is gauge covariant under such transformation\cite{Aris2018PRB} provided that the same wave function is used at the initial point and the final point, i.e. $| u(\boldsymbol{k}_{N+1}) \rangle=| u(\boldsymbol{k}_1) \rangle$. Here we use the same way to show our numerical formula inherits this property so it's appropriate for calculation.

First, covariant derivatives transform as states under the $U(D)$ gauge transformation
\begin{equation}
\begin{aligned}
    \left|\overline{u}_{n, \boldsymbol{k}+\boldsymbol{q}}\right\rangle &\rightarrow \sum_{n^{\prime}}\left(U(\boldsymbol{k})^{\dagger} S_{\boldsymbol{k}, \boldsymbol{k}+\boldsymbol{q}} U(\boldsymbol{k}+\boldsymbol{q})\right)^{-1}_{n^{\prime} n}\left|u_{n^{\prime}, \boldsymbol{k}+\boldsymbol{q}}\right\rangle \\
    &= \sum_{n^{\prime},m,l,m^{\prime}}U(\boldsymbol{k}+\boldsymbol{q})^{-1}_{n^{\prime}m} (S_{\boldsymbol{k}, \boldsymbol{k}+\boldsymbol{q}}^{-1})_{m l} U(\boldsymbol{k})_{l n} \\ & \qquad \qquad U(\boldsymbol{k}+\boldsymbol{q})_{m^{\prime}n^{\prime}} \left|u_{m^{\prime}, \boldsymbol{k}+\boldsymbol{q}}\right\rangle \\
    &= \sum_{m,l} (S_{\boldsymbol{k}, \boldsymbol{k}+\boldsymbol{q}}^{-1})_{m l} U(\boldsymbol{k})_{l n} \left|u_{m, \boldsymbol{k}+\boldsymbol{q}}\right\rangle \\
    &= \sum_{l} U(\boldsymbol{k})_{l n} \left|\overline{u}_{l, \boldsymbol{k}+\boldsymbol{q}}\right\rangle
\end{aligned}
\end{equation}
which makes the covariant derivative expression of Roth term \eqref{Roth form exp2} transform covariantly. This is also true for the matrix elements expression of Roth term \eqref{Roth form exp1} and spin matrix $\sigma_z$, meaning that
\begin{equation}
\begin{aligned}
    \boldsymbol{\mathrm{R}}(\boldsymbol{k}_i)_{m n} \cdot d \boldsymbol{k}_i
    &\rightarrow U(\boldsymbol{k}_{i})^{-1} \boldsymbol{\mathrm{R}}(\boldsymbol{k}_i)_{m n} \cdot d \boldsymbol{k}_i U(\boldsymbol{k}_i) \\
    \sigma_z(\boldsymbol{k}_i)_{m n} \cdot d \boldsymbol{k}_i
    &\rightarrow U(\boldsymbol{k}_{i})^{-1} \sigma_z(\boldsymbol{k}_i)_{m n} \cdot d \boldsymbol{k}_i U(\boldsymbol{k}_i)
\end{aligned}
\end{equation}

Therefore, the second term in \eqref{discretization} is also gauge covariant

\begin{equation}
\begin{aligned}
    &\mathrm{exp}\left[i\boldsymbol{\mathrm{R}}(\boldsymbol{k}_i) \cdot d\boldsymbol{k_i}+i Z\frac{\sigma^{z}}{v^{\perp}}|d\boldsymbol{k}_i|\right] \\ &\rightarrow \mathrm{exp}\left\{i U(\boldsymbol{k}_i)^{-1}\left[\boldsymbol{\mathrm{R}}(\boldsymbol{k}_i) \cdot d\boldsymbol{k_i}+Z\frac{\sigma^{z}}{v^{\perp}}|d\boldsymbol{k}_i|\right]U(\boldsymbol{k}_i)\right\} \\
    &= U(\boldsymbol{k}_i)^{-1}\mathrm{exp}\left[i\boldsymbol{\mathrm{R}}(\boldsymbol{k}_i) \cdot d\boldsymbol{k_i}+i Z\frac{\sigma^{z}}{v^{\perp}}|d\boldsymbol{k}_i|\right] U(\boldsymbol{k}_i)
\end{aligned}
\label{covariance R}
\end{equation}

Besides, the overlap matrix $M^{i}_{mn}=\left\langle u_{m \boldsymbol{k_{i+1}}}|u_{n \boldsymbol{k_i}}\right\rangle$ transforms like

\begin{equation}
\begin{aligned}
    M^{i}
    \rightarrow U(\boldsymbol{k}_{i+1})^{-1} M^{i} U(\boldsymbol{k}_i)
\end{aligned}
\end{equation}

Hence, the covariance of discretized propagator \eqref{discretization} follows from the transformation properties of the two separate terms as

% \begin{widetext}
\begin{eqnarray}
\begin{aligned}
    \mathcal{A}[\mathfrak{o}] &\rightarrow \prod_{i=1}^{N} U(\boldsymbol{k}_{i+1})^{-1} M^{i} U(\boldsymbol{k}_i) \cdot U(\boldsymbol{k}_i)^{-1} \\ &\qquad \mathrm{exp}\left[i\boldsymbol{\mathrm{R}}(\boldsymbol{k}_i) \cdot  d\boldsymbol{k_i}+i Z\frac{\sigma^{z}}{v^{\perp}}|d\boldsymbol{k}_i|\right]U(\boldsymbol{k}_i) \\
    &= U(\boldsymbol{k}_{N+1})^{-1} \\ &\left\{\prod_{i=1}^{N} M^{i} \cdot \mathrm{exp}\left[i\boldsymbol{\mathrm{R}}(\boldsymbol{k}_i) \cdot  d\boldsymbol{k_i}+i Z\frac{\sigma^{z}}{v^{\perp}}|d\boldsymbol{k}_i|\right]\right\} U(\boldsymbol{k}_1) \\
    &= U(\boldsymbol{k}_1)^{-1} \mathcal{A}[\mathfrak{o}] U(\boldsymbol{k}_1)
\end{aligned}
\end{eqnarray}
% \end{widetext}

Since propagator $\mathcal{A}[\mathfrak{o}]$ transforms covariantly, its spectrum $\{e^{i\lambda_a}\}_{a=1}^{D}$ is gauge invariant. In other words, the phase $\lambda_a$ obtained through these numerical formulas is uniquely determined (module $2\pi$) independent of gauge choice in the calculation.

\bibliography{Reference}

\end{document}